\documentclass[prl,reprint, nofootinbib,amsmath,amssymb,aps,floatfix]{revtex4-1}
\usepackage{amsthm}
\usepackage[shortlabels]{enumitem}
\usepackage{braket}
\usepackage{bbold}
\usepackage{comment} 
\usepackage{graphicx, xcolor, varwidth}
\usepackage{tikz,tikz-3dplot}
\usepackage{color}
\usepackage{mathrsfs}

\DeclareMathOperator{\cl }{cl}
\DeclareMathOperator{\setint }{int}
\DeclareMathOperator{\A}{Area}

\renewcommand{\S}{S_{\rm gen}}
\newcommand{\hmax}{H_{\rm max}}

\newcommand{\ch}{\mathcal{C}}
\newcommand{\scri}{\mathscr{I}}

\newcommand{\hmg}{H_{\rm max,gen}}

\begin{document}
\title{Robust Singularity Theorem}
\author{Raphael Bousso}
\affiliation{Center for Theoretical Physics and Department of Physics,\\
University of California, Berkeley, California 94720, U.S.A. 
} 
\begin{abstract}
We prove the Penrose-Wall singularity theorem in the full semiclassical gravity regime, significantly expanding its range of validity. To accomplish this, we modify the definition of quantum-trapped surfaces without affecting their genericity. Our theorem excludes controlled ``bounces'' in the interior of a black hole and in a large class of cosmologies.
\end{abstract}
\maketitle

\paragraph{Introduction}

Penrose's 1965 singularity theorem~\cite{Penrose:1964wq} is of central importance to our understanding of gravity: in generic settings, General Relativity predicts its own breakdown. However, the theorem assumes that the stress tensor twice contracted with any null vector is nowhere negative: the Null Energy Condition (NEC). This condition is violated by valid quantum states in any relativistic field theory. Moreover, it is difficult to exclude the possibility that such violations arise in generic gravitational collapse. This undermines the theorem.

The situation is reminiscent of Hawking's area theorem~\cite{Hawking:1971tu}, which also requires the NEC. Ironically, the NEC \emph{always} fails near a generic quiescent black hole horizon: black holes evaporate~\cite{Hawking:1974sw} and their area decreases, in violation of Hawking's theorem. However, a quantum-corrected area theorem, better known as the Generalized Second Law of Thermodynamics (GSL)~\cite{Bekenstein:1972tm}, appears to survive: the generalized entropy $\S=A/4G\hbar + S_{\rm out}$, the sum of horizon entropy and the matter entropy in the black hole exterior, does not decrease.  

The validity of the GSL in semiclassical gravity is conjectural. No counterexample has been found. Moreover, the GSL follows from well-tested conjectures such as the Bousso Bound~\cite{Bousso:1999xy,Flanagan:1999jp}, which in turn follows from the Quantum Focusing Conjecture (QFC)~\cite{Bousso:2015wca}. The QFC plays a central role in proofs of the properties of entanglement wedges~\cite{Akers:2016ugt,Bousso:2023sya}; and it led to the discovery of the Quantum Null Energy Condition (QNEC), a lower bound on the stress tensor that was later proven directly in quantum field theory~\cite{Bousso:2015mna,Balakrishnan:2017bjg,Ceyhan:2018zfg}. We view these results as overwhelming evidence that the GSL does hold.

Based on just a fraction of this evidence, Wall~\cite{Wall:2010jtc} proposed a profound change of perspective on the Penrose theorem. Wall eliminated the NEC in favor of assuming the GSL, and introduced the novel notion of quantum-trapped surfaces as a diagnostic of singularities.\footnote{Below, ``quantum'' will be omitted before ``trapped'' or ``expansion''; when we mean a classical quantity, we will spell this out.} This moved the prediction of singularities onto a new, stronger foundation. However, semiclassical gravity -- the natural setting of the Penrose-Wall theorem -- is far broader than the narrow limit in which its proof is valid.

In the present paper, we present a significant expansion of the Penrose-Wall singularity theorem. Our result excludes ``bounces'' or any other form of singularity avoidance inside black holes, so long as the semiclassical regime remains valid through the bounce.\footnote{It is difficult to think of any other regime in which a bounce could be operationally established. Note that causal structure would have to be preserved in order to transmit information through a bounce and thus make the bounce verifiable. But then the GSL is applicable across the bounce and the contradiction established in our theorem still excludes the bounce.} It similarly excludes ``bounces'' in cosmology with noncompact Cauchy slices, such as pre-big-bang~\cite{Veneziano:2000pz} and cyclic~\cite{Khoury:2001wf,Graham:2017hfr} models. (Bounces with compact Cauchy slices can be excluded by singularity theorems based on entropy bounds~\cite{Bousso:2022cun,Bousso:2022tdb}, whose extension to the full semiclassical regime will be presented in future work.)

\paragraph{Semiclassical gravity as an asymptotic expansion}

A narrow view of semiclassical gravity would define it as a perturbative expansion in $G\hbar$ in the strict limit as $G\hbar\to 0$ (more precisely, the limit as $G\hbar/A\to 0$, where $A$ is any other quantity of dimension area). This expansion can be invoked to solve the Einstein equation $G_{ab}=8\pi G \braket{T_{ab}}$ iteratively in powers of $G\hbar$, or to justify a saddlepoint approximation to the Euclidean gravitational path integral. For example, in a universe dominated by pressureless dust, the stress tensor can be treated as a classical fluid at leading order ($\hbar^0$); one-loop corrections to $\braket{T_{ab}}$ will be proportional to $\hbar$ and thus will become arbitrarily small if the $G\hbar$ limit is implemented by taking $\hbar\to 0$ at fixed $G$. 

While the proof of the Penrose-Wall theorem
applies only in this narrow limit, semiclassical gravity has been successfully applied -- and experimentally tested -- in a far broader regime: a strict limit as $G\hbar\to 0$ is not required. For example, the stress tensor in our own universe was dominated by thermal radiation until it was about 30,000 years old. The radiation epoch is tightly constrained observationally, and all known data can be understood accurately from semiclassical gravity and the Standard Model of particle physics. (This includes the era of Big Bang Nucleosynthesis, from which the abundances of light elements can be successfully predicted with great precision.) In the radiation epoch, the stress tensor $\braket{T_{ab}}\sim \hbar$ cannot be approximated as classical without changing the temperature of the radiation. The Hubble horizon area scales as $(G\braket{T_{ab}})^{-1}\sim (G\hbar)^{-1}$, so the $G\hbar\to 0$ limit cannot be taken without changing observable quantities. A narrow definition of the semiclassical regime that invokes this limit would thus be inapplicable to a well-studied era of our own universe.

Significant theoretical advances would also be eliminated if semiclassical gravity became invalid at finite $G\hbar$. There would be no basis for the conclusion that Hawking radiation causes a black hole to shrink by any finite fraction of its initial size, since the evaporation time scales as $(G\hbar)^{-1}$. The GSL and the Bousso bound would be reduced to their field theory limits as $G\hbar\to 0$, where they remain nontrivial and become provable~\cite{Casini:2008cr,Wall:2011hj,Bousso:2014sda,Bousso:2014uxa} but are greatly diminished in scope. Similarly, the QFC would reduce entirely to the QNEC. The recent semiclassical~\cite{Lewkowycz:2013nqa} derivation~\cite{Penington:2019npb,Almheiri:2019psf} of the microscopic Hawking radiation entropy from the Quantum Extremal Surface prescription~\cite{Ryu:2006bv,Hubeny:2007xt,Faulkner:2013ana,Engelhardt:2014gca} also requires finite $G\hbar$ and thus would not survive a strict $G\hbar\to 0$ limit.

Instead, semi-classical gravity is best characterized as an asymptotic expansion at small \emph{finite} $G\hbar$. This viewpoint allows us to include the radiation epoch, Hawking evaporation, and entanglement islands within the arena of semiclassical gravity; and it preserves the full power of the GSL, the Bousso bound, and the QFC. This regime can be defined most precisely in settings with a large number $c$ of light matter fields, where the small finite asymptotic expansion parameter becomes $cG\hbar$. In order to suppress metric fluctuations, the limit $G\hbar\to 0$ can be taken at fixed $cG\hbar$. This preserves dynamical gravity, including full black hole evaporation. A concrete example is the brane world scenario~\cite{Randall:1999vf}, where $c$ is the central charge of a CFT on the brane, and $cG\hbar$ is a power of the curvature radius of the bulk. In this setting, a version of the QFC, and hence the Bousso bound and GSL, can be explicitly proven through a classical bulk computation that is \emph{exact} in $cG\hbar$~\cite{Shahbazi-Moghaddam:2022hbw}. (I thank A.~Shahbazi-Moghaddam for emphasizing the significance of this result to me.) By a related method, the Page curve calculation~\cite{Penington:2019npb,Almheiri:2019psf} can be proven~\cite{Almheiri:2019hni}. However, a key step in the proof of the Penrose-Wall theorem fails at finite $cG\hbar$.

\paragraph{Touching Lemma and Its Limitations}

\begin{figure}
    \centering
    \includegraphics[width=0.8\linewidth]{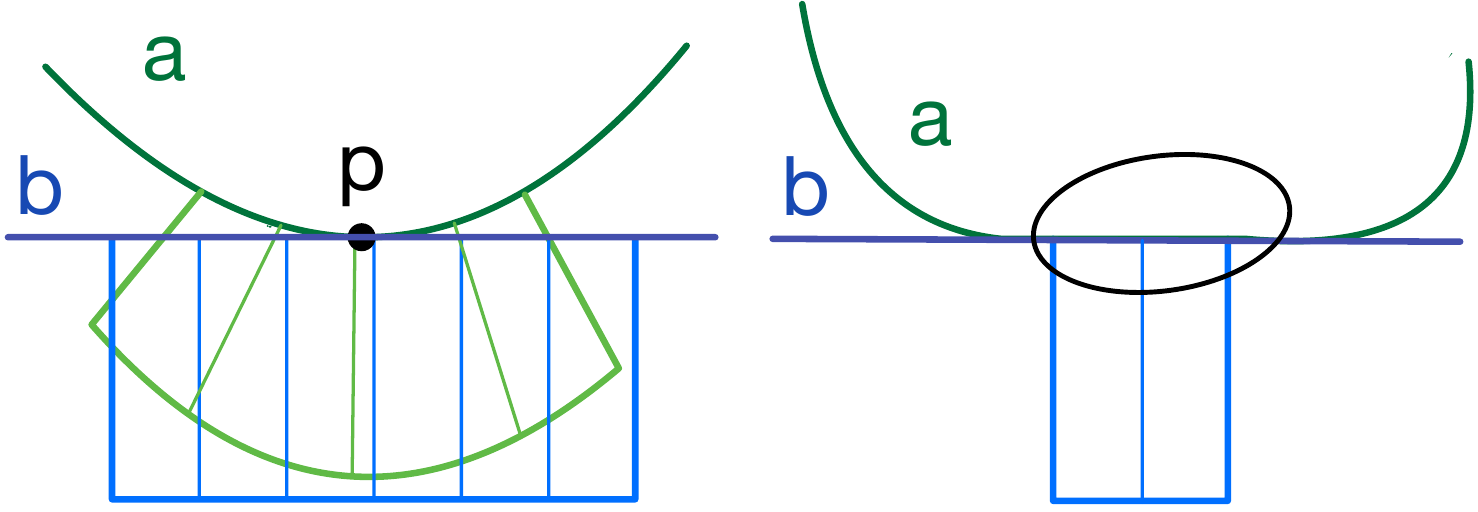}
    \caption{Spatial projection onto a Cauchy slice. The edges of the regions $a\subset b$ touch at the point $p$ (left figure). Classically, the outward expansion of $a$ is obviously no smaller than that of $b$ at $p$. But the difference in quantum expansions is unconstrained, because a null deformation of $a$ at $p$ (light green) will necessarily access a different region than any null deformation of $b$ at $p$ (light blue) -- except if the edges of $a$ and $b$ overlap in an entire open set (black ellipse, right figure). Only the latter situation arises in the new proof given here.}
    \label{fig:touching}
\end{figure}

The step that confines Wall's proof~\cite{Wall:2010jtc} to the strict $cG\hbar\to 0$ limit is the ``touching lemma,'' which provides an inequality between the expansions of two nested surfaces that touch but do not agree on any open neighborhood. At the classical level the lemma is obvious; see Fig.~\ref{fig:touching}. 
When area is replaced by $\S$, the expansion must be defined as a functional derivative of $\S$ under shape deformations of each surface~\cite{Wall:2010jtc}. The orthogonal null geodesics that generate deformations of two touching surfaces access different new spacetime regions; see Fig.~\ref{fig:touching}. This leads to a loss of control in attempting to establish an inequality. As a workaround, one can argue~\cite{Wall:2010jtc} that \emph{in the strict $cG\hbar\to 0$ limit}, the classical expansions will dominate, unless the two surfaces agree on a whole open neighborhood $O$. If they do not, then the proof reduces to the trivial classical case. If they do, the shape deformations can be restricted to $O$ so that they are identical for both surfaces; the desired inequality then follows from the strong subadditivity of the von Neumann entropy $S_{\rm out}$~\cite{Wall:2010jtc}. 

\begin{figure}
    \centering
    \includegraphics[width=0.7\linewidth]{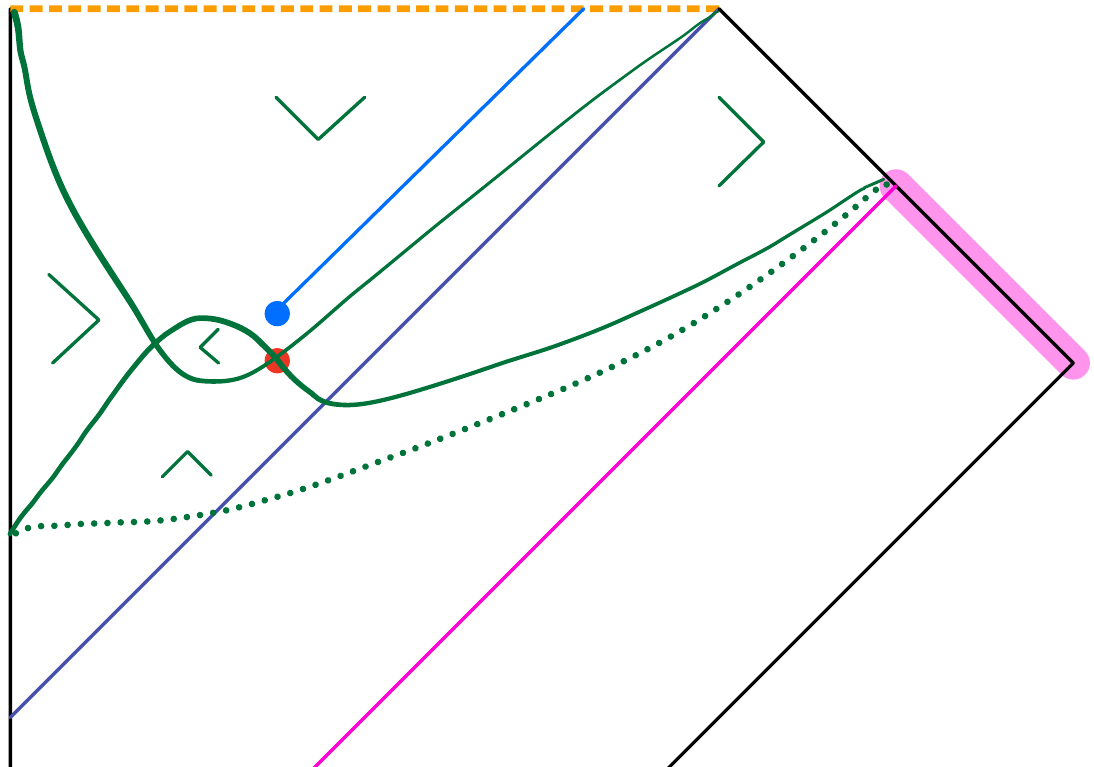}
    \caption{Penrose diagram of an evaporating black hole; the event horizon is blue and the future singularity is orange. The nonexpanding null directions of spheres are shown in green~\cite{Bousso:1999cb}. (For the green dotted line see Ref.~\cite{Bousso:2022tdb}.) The pink lines denote the Hawking radiation that has arrived at $\scri^+$ (thick) and its causal past (thin). The red dot marks the Quantum Extremal Surface associated with the radiation, at which the classical null expansions are canceled by an equal and opposite quantum term. The two terms remain comparable at the sphere marked by the blue dot; therefore a strict $cG\hbar \to 0$ limit cannot be taken while holding the overall geometry fixed. The blue sphere is quantum trapped, even after it is slightly perturbed to make its shape generic. The blue geodesic is incomplete, but the proof of the Penrose-Wall singularity theorem does not apply because it requires $cG\hbar\to 0$. Here we prove a theorem that applies at fixed nonzero $cG\hbar$.}
    \label{fig:nontouching}
\end{figure}

Without taking a $cG\hbar\to 0$ limit, this argument fails, since the quantum portion of the expansion need not be negligible. ``Entanglement islands''~\cite{Penington:2019npb,Almheiri:2019psf} furnish an explicit example of a controlled semiclassical calculation in which the classical and quantum contributions to the expansion are by construction \emph{equal} in magnitude. Near the QES lie quantum-trapped surfaces for which the conclusion of the Penrose-Wall theorem holds but its proof does not apply; see Fig.~\ref{fig:nontouching}.
In order to prove a singularity theorem that applies in this regime, we will need to sidestep the touching lemma.

\paragraph{Why Wedges? Why Max Entropies? Why Discrete?}

Our formalism invokes several concepts that did not appear in Ref.~\cite{Wall:2010jtc}: wedges instead of partial Cauchy slices; the conditional max entropy instead of von Neumann entropy; and discrete notions of (non-)expansion or contraction instead of numerical values. The present work is self-contained; but the reader familiar with the older literature may be interested in the origins of these differences. We will review them briefly while omitting many details that can be found in the references.

Causal development of a spatial region, whether by classical Hamiltonian evolution or by unitary quantum evolution, preserves its information content. Thus one should associate any information measure not to a specific partial Cauchy slice but to its full domain of dependence. This domain is called a wedge and defined below.

Our choice of information measure is dictated by the  recent demonstration that the conditional max entropy rather than the von Neumann entropy \emph{must} be used for constructing entanglement wedges~\cite{Akers:2020pmf}. To avoid a proliferation of conjectures (such as a max, min, and von Neumann QFC~\cite{Akers:2023fqr}), one would then like to use \emph{only} max-entropies (and consider only outward null deformations of wedges, since min- and max-entropies are exchanged under wedge complementarity). Fortunately, the max-QFC is indeed sufficient for deriving all important properties of entanglement wedges~\cite{Bousso:2024iry}; similarly, the max-GSL (which is a consequence of the max-QFC) will suffice here. The discrete (signed) definitions of expansion, too, maximize generality while retaining key results~\cite{Shahbazi-Moghaddam:2022hbw,Bousso:2024iry}. 

None of these updates are critical to our proof, so one may freely substitute partial Cauchy slices for wedges, von Neumann for max entropies, and numerical quantum expansions for the discrete notions of past-nonexpanding (PNE) and past-noncontracting (PNC) defined below. What is critical to generalizing the Penrose-Wall singularity theorem is the notion of PNC itself, which enters the new definition ``trapped'' we give below: it is inequivalent to the definition used by Wall~\cite{Wall:2010jtc} even at the level of numerical von Neumann quantum expansions of partial Cauchy slices. 

This new definition of ``trapped'' is, however, the only one we were able to construct from the bare-bones structure of discrete max expansions developed in Ref.~\cite{Bousso:2024iry}. In that sense, the more general theorem presented here did grow out of recent developments.

\paragraph{Causal Structure and Wedges}
\label{sec:preliminarydefinitions}

Let $(M,g)$ be an inextendible globally hyperbolic spacetime. A \emph{curve} is a continuous map from an interval $C\subset \mathbb{R}$ into $M$. A curve is \emph{timelike (causal)} if its tangent vector is everywhere timelike (non-spacelike). 
For any set $s\subset M$, we denote its interior by $\setint s$, its closure by $\cl s$, and its boundary in $M$ by $\partial s$. The chronological future of $s$, $I^+(s)$, is the set of points $q$ such that there exists a future-directed timelike curve that begins at some point $p\in s$ and ends at $q$. $I^-(s)$ is defined analogously, and $I(s)\equiv I^+(s)\cup I^-(s)$. (We omit further duplication of past/future definitions.)

A null geodesic is called \emph{future-complete} if it can be extended to infinite affine parameter into the future; otherwise, it is called future-incomplete. $(M,g)$ is called \emph{null-geodesically incomplete} -- and hence, singular~\cite{Penrose:1964wq} -- if it contains at least one future- or past-incomplete null geodesic. (Definitions so far follow Wald~\cite{Wald:1984rg}.)

Following Refs.~\cite{Bousso:2023sya,Bousso:2024iry} (where more details and figures are given), the \emph{spacelike complement} of a set $s\subset M$ is $s'\equiv \setint [M\setminus I(s)]$. A {\em wedge} is a set $a\subset M$ that satisfies $a=a''$. (Thus a wedge is open, and the intersection of two wedges is a wedge.) The {\em wedge union} of two wedges $a,b$ is the wedge $a\Cup b\equiv (a'\cap b')'$. Given a wedge $a$, its \emph{edge} is $\eth a \equiv \partial a \setminus I(a)$ and its \emph{future Cauchy horizon} is  $H^+(a) \equiv \partial a\cap I^+(a)$. The area of $\eth a$ will be denoted $\A(a)$. The following properties follow immediately: 
$H^+(a)$ is a Cauchy slice of $a$; $\eth a$ and $\eth a \cup \Sigma_a$ are closed in $M$; and $a = \Sigma_a''$, where $\Sigma_a$ is any Cauchy slice of $a$.

Let the {\emph unphysical spacetime} $(\tilde M,\tilde g)$  be the conformal completion~\cite{Wald:1984rg} of $(M,g)$ by addition of a \emph{conformal boundary}, $\partial M\equiv \tilde M\setminus M$. \emph{Future null infinity}, $\scri^+$, is the subset of $\partial M$ consisting of the future endpoints in $\tilde M$ of null geodesics of future-infinite affine length in $M$. 

A \emph{causal horizon} $\ch$ is the boundary of the past of any subset $s\subset \scri^+$ of null infinity: $\ch= \dot I^-(s).$ (In many common examples such as black hole horizons, Rindler horizons, and the cosmological horizon of de Sitter space, $s$ consists of a single point; but this is not assumed in our definition.) We will call a wedge $a$ the \emph{exterior} of $\ch$ at the \emph{cut} $\eth a$ if $\eth a\subset \ch$ and $I^-(s)$ contains a Cauchy slice of $a$. (Note that not all Cauchy slices of $a$ need to be in $I^-(s)$; for example, if $\ch$ is the intersection of two Rindler horizons in Minkowski space, then at sufficiently early times the exterior of $\ch$ is $M$.) Given a Cauchy slice $\Sigma$ of $M$, the exterior of $\ch$ with cut $\Sigma\cap \ch$ is the wedge $[\Sigma\cap I^-(s)]''$. 
For two exteriors $a$ and $b$ of $\ch$, $a\supset b$ if and only if $\eth a$ is nowhere to the future of $\eth b$.

\paragraph{Generalized Max Conditional Entropy} For any nested pair of wedges $a\subset b\subset M$, there exists a family, parametrized by $\epsilon$, of \emph{generalized smooth conditional max entropies}, $\hmg^\epsilon(b|a)$. To leading order in $G\hbar$ they are defined by
\begin{equation}\label{eq:hmgexpansion}
    \hmg^\epsilon(b|a) = \frac{\A(b)-\A(a)}{4G\hbar} + \hmax^\epsilon(b|a) + O(G\hbar)~,
\end{equation}
where $\hmax^\epsilon(b|a)$ is the standard smooth max entropy of the matter field quantum state restricted to $b$ conditioned on $a$~\cite{RenWol04a,Akers:2020pmf,Akers:2023fqr}. 
Each right hand side term in Eq.~\eqref{eq:hmgexpansion} requires a short-distance cutoff on which both Newton's constant and $\hmax(b|a)$ will depend. There is strong evidence~\cite{Bousso:2015mna} that the sum $\hmg^\epsilon(b|a)$ is independent of the cutoff. Moreover, a more careful algebraic definition of $\hmg$ allows for fluctuations of the area by treating it as an operator~\cite{Akers:2023fqr}; thus the above expansion is included mainly as a heuristic. As noted above, a reader unfamiliar with single-shot entropies can replace $\hmg$ by the conditional generalized entropy, $\hmg(b|a)\to \S(b|a) = \S(b)-\S(a) = \A(b)-\A(a)+S_{\rm out}(b)-S_{\rm out}(a)$,
so long as the relevant bulk quantum states are compressible~\cite{Akers:2020pmf}. We shall work directly with $\hmg^\epsilon$; to avoid clutter, we will drop the $\epsilon$ below.

$\hmg$ satisfies \emph{strong subadditivity}: let $a\subset b$ be wedges. Then for any wedge $c$ spacelike to $b$,
\begin{equation}
        \hmg(b\Cup c|a\Cup c)\leq \hmg(b|a)~.
\end{equation}
In fact, strong subadditivity holds separately for the area and for $\hmax(b|a)$. At higher orders in $G\hbar$, it should be considered a conjecture~\cite{Bousso:2024iry}.

\paragraph{Discrete Max-Expansions} A wedge $a$ is called \textit{past-nonexpanding} (PNE) at $p \in \eth a$ if there exists an open set $O$ containing $p$ such that $\hmg(b|a) \leq 0 \text{ for all wedges } b \supset a \text{ such that } \eth b \subset \eth a \cup [H^-(a') \cap O]$.
A wedge $a$ is \emph{past-noncontracting (PNC)} if there exists an open set $O \supset \eth a$ such that no proper future-directed outward null deformation of $a$ within $O$ is PNE at all new edge points. That is, no wedge $b \supsetneq a$ with $\eth b \subset \eth a \cup [H^+(a') \cap O]$ is PNE on all points in $\eth b\setminus \eth a$. 

\paragraph{\textbf{Conjecture (Generalized Second Law of Thermodynamics)}}
    Let $s\subset \scri^+$ and let $\ch =\dot I^-(s)$ be a causal horizon with exteriors $a$ and $b$. If $a\supset b$, then $\hmg(a|b)\leq 0$.

\paragraph{\textbf{Lemma}} The GSL immediately implies that the exterior of $\ch$ at any cut is PNE.

\paragraph{\textbf{Definition}} 
    A wedge $a$ is called \emph{trapped} if $\eth a$ is compact and $a$ is past-noncontracting. (This definition deviates substantially from Refs.~\cite{Penrose:1964wq,Wall:2010jtc,Wald:1984rg} even at the classical level. For example, a  cut of a stationary black hole event horizon is considered trapped with our definition. The difference is critical to our ability to prove a singularity theorem in the full semiclassical regime. However, it is easy to verify that generic surfaces that are trapped in the sense of Refs.~\cite{Penrose:1964wq,Wall:2010jtc,Wald:1984rg} are trapped also by the present definition.)

\begin{figure}
    \centering
    \includegraphics[width=0.7\linewidth]{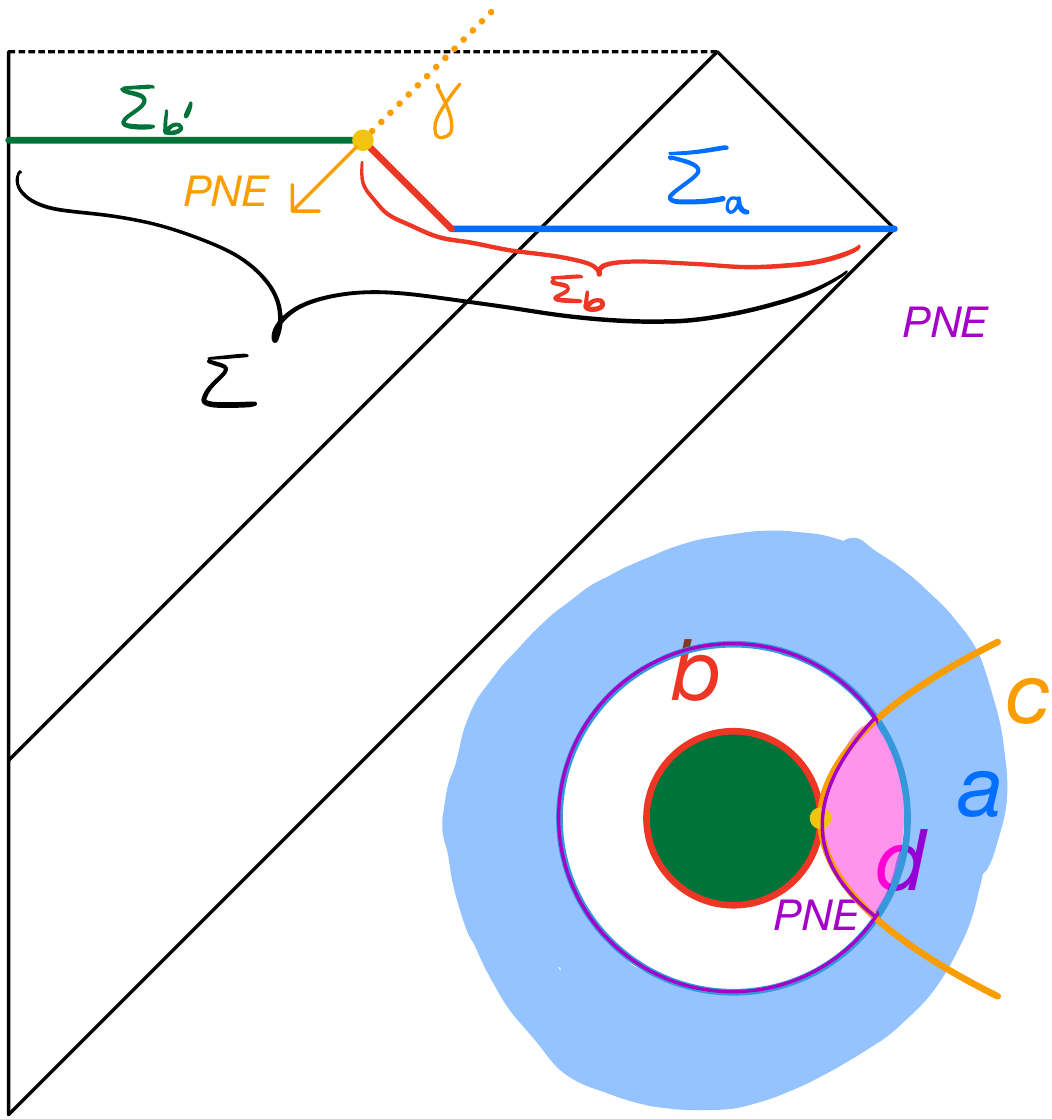}
    \caption{Wedges involved in the proof of the singularity theorem. Top left: conformal spacetime diagram. Bottom right: only the Cauchy slice $\Sigma$ and its subregions are shown. $a$ is the blue shaded region, assumed to be noncompact and trapped (PNC). Its future null deformation $b$ is the exterior of the red circle. Assuming geodesic completeness (no singularity), the null geodesic $\gamma$ defines a causal horizon $\dot I^-(\gamma)$. The orange curve is $\dot I^-(\gamma)\cap \Sigma$, and the wedge $c$ lies to its right. The wedge $d=a\Cup c$ (blue and pink regions combined) must be PNE on the edge portion $\eth d\setminus \eth a$ (the arc where the orange and purple curves coincide), in contradiction with $a$ being PNC.}
    \label{fig:prooffigure}
\end{figure}

\paragraph{\textbf{Semiclassical Singularity Theorem}}

Let $(M,g)$ be an inextendible, globally hyperbolic spacetime containing a trapped wedge $a\subset M$ with Cauchy slice $\Sigma_a$. If\, $\Sigma_a\cup \eth a$ is noncompact, then $M$ is singular. Specifically, there exists an open set $O\supset \eth a$ such that every proper future null outward deformation $b$ of $a$ within $O$
has the property that at least one null geodesic generator of $H^+(b)$ is future-incomplete.

\begin{proof}
    Let $O$ be the open set guaranteed by the assumption that $a$ is PNC. 
    Let $b\supsetneq a$ with $\eth b \subset \eth a \cup [H^+(a') \cap O]$. We note that $\Sigma_b = [\Sigma_a \cup \eth a \cup H^+(a')] \cap b$
    is a Cauchy slice of $b$; see Fig.~\ref{fig:prooffigure}. We will suppose for contradiction that every null geodesic generator of $H^+(b)$ is future-complete. 
    
    At least one null geodesic generator $\gamma$ of $H^+(b)$ must remain on $H^+(b)$ for infinite affine time. Otherwise, $H^+(b)\cup \eth b$ would be compact, so that $\Sigma_b\cup \eth b$ would be compact~\cite{Penrose:1964wq,Wall:2010jtc}, and hence its closed subset $\Sigma_a\cup \eth a$ would also be compact, in contradiction with its assumed noncompactness.  

    Let $s$ be the endpoint of $\gamma$ on future null infinity, and let $\ch=\dot I^-(s)=\dot I^-(\gamma)$ be the associated causal horizon. Let $\Sigma_{b'}$ be a Cauchy slice of $b'$, so that $\Sigma\equiv \Sigma_{b'}\cup \eth b \cup \Sigma_b$ is a Cauchy slice of $M$. Let $c\equiv [I^-(s)\cap \Sigma]''$ be the exterior of $\ch$ at the cut $\ch\cap \Sigma$. The above Lemma dictates that $c$ is PNE.
    
    Since $s\subset H^+(b)$ in $\tilde M$, no timelike curve connects $\eth b\cup \Sigma_{b'}$ to $s$. Hence $\eth c\subset \eth b \cup \Sigma_b\subset O$. Thus $c\subset b$, and $c$ admits the Cauchy slice
    \begin{equation}
        \Sigma_c=\Sigma_b\cap c = [\Sigma_a \cup \partial a \cup H^+(a')]\cap c~.
    \end{equation}
    Let $d=a\Cup c$. Thus $d\supset a$, $\eth d \subset O$, and $d$ admits the Cauchy slice 
    \begin{equation}
        \Sigma_d = [\Sigma_a \cup \eth a \cup H^+(a')]\cap d~.
    \end{equation}
    Moreover, $d\neq a$ since the past endpoint of $\gamma$ on $\eth c$ lies is $\eth d\setminus \eth a$. Hence $d\supsetneq a$ and $\eth d \subset \eth a \cup [H^+(a') \cap O]$, so $d$ is a proper future outward deformation of $a$ within $O$.
    
    By construction, $\eth c$ and $\eth a$ are nowhere timelike related, so $\eth d\subset \eth c \cup \eth a$ and hence $\eth d\setminus \eth a \subset \eth c$. Because $c$ is PNE, strong subadditivity of $\hmg$ implies that $d$ is PNE on all points in $\eth d\setminus \eth a$. This contradicts the assumption that $a$ is PNC. Therefore, at least one null geodesic generator of $H^+(b)$ is future-incomplete.
\end{proof}

\paragraph{Acknowledgements}
I would like to thank L.~Bariuan, S.~Kaya, P.~Rath, A.~Shahbazi-Moghaddam, E.~Tabor, A.~Wall and E.~Witten for discussions. I am particularly grateful to A.~Shahbazi-Moghaddam for emphasizing to me the limitations of the touching lemma.

\bibliographystyle{JHEP}
\bibliography{covariant}

\end{document}